\input{aipcheck}

\documentclass[
    ,final            
  ]
  {aipproc}

\layoutstyle{6x9}

\begin{document}
\newcommand{\be}{\begin{equation}}
\newcommand{\ee}{\end{equation}}
\newcommand{\bea}{\begin{eqnarray}}
\newcommand{\eea}{\end{eqnarray}}
\newcommand{\bT}{{${\bf b}_\perp$}}
\newcommand{\xT}{{${\bf x}_\perp$}}
\newcommand{\bi}{\begin{itemize}}
\newcommand{\ei}{\end{itemize}}
\title{Quark and Gluon Angular Momentum in the Nucleon}


\author{Matthias Burkardt}{
  address={New Mexico State University}
}

\author{Hikmat BC}{
  address={New Mexico State University}
}

\author{Abdullah Jarrah}{
  address={New Mexico State University} 
}

\begin{abstract}
 Parton distributions in impact parameter space, which
are obtained by Fourier transforming GPDs, are exhibit
a significant deviation from axial symmetry when the
target and/or quark is transversely polarized. From this
deformation, we present an intuitive derivation of the Ji
relation. 
In a scalar diquark model and in QED, we compare the Ji and Jaffe-Manohar
decompositions of the nucleon spin. Using the MIT bag model,
we estimate spectator effects through the presence of the gluon vector potential in the definitions of the quark orbital angular momentum.
\end{abstract}

\maketitle


\section{Distribution of Quarks in the Transverse Plane}

\begin{figure}
\unitlength1.cm
\begin{picture}(10,5)(2.7,12.7)
\includegraphics{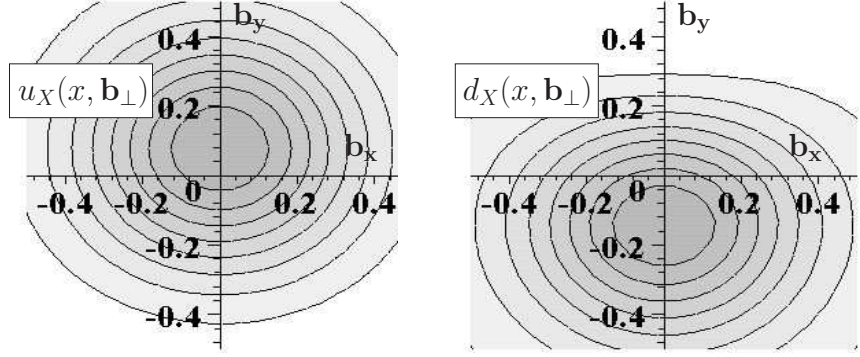}
\end{picture}
\caption{Distribution of the $j^+$ density for
$u$ and $d$ quarks in the
$\perp$ plane ($x=0.3$ is fixed) for a proton that is 
polarized
in the $x$ direction in the model from
Ref. \cite{IJMPA}.
For other values of $x$ the distortion looks similar.
}
\label{fig:distort}
\end{figure}

In the case of transversely polarized quarks and/or nucleons, parton
distributions in impact parameter space \cite{mb1}
show a significant 
transverse deformation. In the case of unpolarized quarks in a nucleon
polarized in the $+\hat{x}$ direction, this deformation is described
by the $\perp$ gradient of the Fourier transform of the GPD $E^q$ 
\cite{IJMPA}
\be
q_{q/p\uparrow}(x,{\bf b}_\perp) = \int \frac{d^2{\bf x}_\perp}{(2\pi)^2}
e^{-i{\bf b}_\perp \cdot { \Delta}_\perp} 
H^q(x,0,-\Delta_\perp^2) - 
\frac{1}{2M} \partial_y \int \frac{d^2{\bf x}_\perp}{(2\pi)^2}
e^{-i{\bf b}_\perp \cdot {\Delta}_\perp} 
E^q(x,0,-{ \Delta}_\perp^2)
\label{eq:shift}
\ee
for quarks of flavor $q$. Since $E^q(x,0,t)$ also arises in the 
decomposition of the Pauli form factor $F_2^q=\int_{-1}^1 dx
E^q(x,0,t)$ for quarks with flavor $q$ (here it is always understood
that charge factors have been taken out) w.r.t. $x$, this allows 
to relate the $\perp$ flavor dipole moment to the contribution from
quarks with flavor $q$ to the nucleon anomalous magnetic moment
(here it is always understood
that charge factors have been taken out)
\be
d^q \equiv \int d^2{\bf b}_\perp q_{+\hat{x}}(x,{\bf b}_\perp)
b_y = \frac{1}{2M} F_2^q(0)=\frac{1}{2M} \kappa_{q/p}.
\ee
Here $e_q\kappa_{q/p}$ is the contribution from flavor $q$ to the
anomalous magnetic moment of the proton. Neglecting the contribution
from heavier quarks to the nucleon anomalous magnetic 
moment, one can use the proton and neutron anomalous
magnetic moment to solve for the contributions from $q=u,d$, yielding
$\kappa_{u/p} \approx 1.67$ and $\kappa_{d/p}\approx -2.03$. 
The resulting deformation ($|d_q|\sim 0.1 \mbox{fm}$)
of impact parameter dependent PDFs in the transverse direction 
(fig. \ref{fig:distort}) is rather significant and it is in opposite 
directions for $u$ and $d$ quarks.
The sideways displacement of the center of momentum for each quark 
flavor from the origin provides a very intuitive derivation of the
Ji-relation \cite{JiPRL}
for the contribution from quarks with flavor $q$ to the nucleon 
angular momentum:
Consider nucleon state that is an eigenstate
under rotation about the $\hat{x}$-axis (e.g. a wave packet 
describing a nucleon  polarized in
the $\hat{x}$ direction
with ${\vec p}=0$). 
For such a state, $\langle T_q^{00}y\rangle=0=\langle 
T_q^{zz}y\rangle$ and 
$\langle T_q^{0y}z\rangle=- \langle T_q^{0z}y\rangle$, and
therefore
$\langle T_q^{++}y\rangle = \langle T_q^{0y}z-T_q^{0z}y\rangle
= \langle J^x_q\rangle$.
This result allows to relate the $\perp$ shift of the center of 
momentum for quark flavor $q$
to the angular momentum $J_q^x$ carried by that quark flavor.
The displacement of the $\perp$ center of 
momentum is a sum of two effects:
\bi
\item as discussed above,  $\langle T_q^{++}y\rangle$ for a 
quark relative to
the center of momentum of a transversely polarized nucleon 
(Fig. \ref{fig:distort})
\item however, already for a point-like transversely 
polarized spin $\frac{1}{2}$ particle, the
$\perp$ center of momentum $\langle T^{++}y\rangle$ is shifted
by $\frac{1}{2}$ a Compton wavelength away from the origin
(the center of the wave packet in the rest frame)
\ei
In order to understand the $2^{nd}$ effect, i.e. the $\perp$ shift of the
center of momentum for a $\perp$ polarized spin $\frac{1}{2}$
particle, let us consider `bag model' \cite{MIT} type wave functions\footnote{The precise
shape of the wave packet does not matter as long as its size is sent to $\infty$.} for the 
wave packet of the target, but with
a `bag radius' $R$ that will be sent to $\infty$ at the end
\be
\psi = \left( \begin{array}{c} f(r)\\
\frac{{\vec \sigma}\cdot {\vec p}}{E+m} f(r)\end{array} \right) \chi
\quad\quad \mbox{with} \quad \quad \chi= \frac{1}{\sqrt{2}}
\left( \begin{array}{c}1\\ 1\end{array} \right).
\ee
Since $\psi^\dagger \partial_z\psi$ is even under $y\rightarrow
-y$, $i\bar{q}\gamma^0 \partial^z q$ does not contribute to
$\langle T^{0z}y\rangle =\langle i\bar{q}\left(\gamma^0 \partial^z + \gamma^z 
\partial^0\right)q\rangle$.
Using $i\partial_0\psi=E\psi$, one thus finds
\bea 
\langle T^{0z}b_y\rangle &=& E\int d^3r \psi^\dagger \gamma^0
\gamma^z \psi y = E\int d^3r \psi^\dagger \left(
\begin{array}{cc} 0 & \sigma^z\\ \sigma^z & 0 \end{array}\right)\psi
y
\\
&=& \frac{2E}{E+M}\int d^3r \chi^\dagger \sigma^z\sigma^y \chi
 f(r)(-i)\partial^y f(r) y
= \frac{E}{E+M} \int d^3r f^2(r).\nonumber\eea
In the limit when the bag radius goes to infinity $E=M$ and
$\int d^3r f^2(r)=1$ and therefore the
$2^{nd}$ moment of $\perp$ flavor dipole moment for this wave packet reads
\be \langle T_q^{++}y\rangle=\langle T^{0z}b_y\rangle=\frac{1}{2},\ee
i.e.  for a
a spherically symmetric, delocalized (Dirac-)wave packet with $J_x=\frac{1}{2}$  centered around the origin  
the $\perp$ center of momentum $\frac{1}{M}\langle T_q^{++}b_y\rangle$
is 
\underline{not} at origin, but at $\frac{1}{2M}$! 

This `overall shift' of the nucleon $\perp$ center of momentum
implies a contribution $\frac{1}{2}\langle x_q\rangle$ to $\langle T^{++}y\rangle$
from quarks carrying momentum fraction $\frac{1}{2}\langle x_q\rangle$, which
gives rise to the first term in the Ji relation.
In addition, Eq. (\ref{eq:shift}) implies a shift of the center of momentum for quark flavor
$q$ relative to that of the nucleon, by $\frac{1}{2M}\int dx\,x E(x,0,0)$. Combining these two effects yields the Ji relation
\cite{JiPRL}
\be
J_q= \langle T_q^{++}b_y\rangle = \frac{1}{2}\int dx\, x \left[q(x)+E_q(x,0,0)\right]
\label{eq:Ji}
\ee

Since the famous EMC experiments revealed that only a small fraction
of the nucleon spin is due to quark spins \cite{EMC}, 
there has been a great
interest in `solving the spin puzzle', i.e. in decomposing the
nucleon spin into contributions from quark/gluon spin and
orbital degrees of freedom.
In this effort, the Ji decomposition \cite{JiPRL}
\be
\frac{1}{2}=\frac{1}{2}\sum_q\Delta q + \sum_q { L}_q^z+
J_g^z
\label{eq:JJi}
\ee
appears to be very useful, 
as not only the quark spin contributions $\Delta q$ but also
the quark total angular momenta $J_q \equiv \frac{1}{2}\Delta q + 
{ L}_q^z$ (and by subtracting the spin piece also the
the quark orbital angular momenta $L_q^z$) entering this decomposition
can be accessed experimentally through GPDs.
The terms in (\ref{eq:JJi}) are defined as expectation
values of the corresponding terms in the angular momentum tensor
\be
M^{0xy}= \sum_q \frac{1}{2}q^\dagger \Sigma^zq +
\sum_q q^\dagger \left({\vec r} \times i{\vec D}
\right)^zq
+  
\left[{\vec r} \times \left({\vec E} \times {\vec B}\right)\right]^z
\label{M012}
\ee
in an appropriate nucleon wave packet. Here
$i{\vec D}=i{\vec \partial}-g{\vec A}$ is the gauge-covariant
derivative.
The main advantages of this decomposition are that each term can be 
expressed as the
expectation value of a manifestly gauge invariant
local operator and that  $J_q^z=\frac{1}{2}\Delta q+L_q^z$
can be related to GPDs
(\ref{eq:Ji}) 
and is thus accessible in deeply virtual Compton scattering and
meson production and can also be
calculated in lattice QCD.
However, due to the presence of interactions  
 through the vector potential in the gauge covariant derivative
$L_q^z$ does not have a parton interpretation.

Recent lattice calculations of GPDs surprised in several ways
\cite{lattice}.
First, the light quark orbital angular momentum (OAM) is consistent 
with
$L_u\approx -L_d$, i.e. $L_u+L_d\approx 0$, which
would imply that $J_g \approx \frac{1}{2}\cdot 0.7$ represents
the largest piece in the nucleon spin decomposition. Secondly,
$L_u\approx -0.15$ and $L_u\approx +0.15$ in these calculations, 
i.e. the
opposite signs from what one would expect from many quark
models with relativistic effects, as we will also illustrate in the
following section. While the inclusion of still-omitted
disconnected diagrams may change the sum $L_u+L_d$, it does not
affect the difference $L_u-L_d$. In Ref. \cite{Tony}, it was
pointed out that evolution from a quark model scale of few hundred
MeV to the lattice scale of few GeV might explain the
difference. 

Jaffe and Manohar have proposed an alternative decomposition of the
nucleon spin, which does have a partonic interpretation
\cite{JM}
\be
\frac{1}{2}=\frac{1}{2}\sum_q\Delta q + \sum_q {\cal L}_q^z+
\frac{1}{2}\Delta G + {\cal L}_g^z,
\label{eq:JJM}
\ee
and whose terms are defined as matrix elements of the corresponding
terms in the $+12$ component of the angular momentum tensor
\be
M^{+12} = \frac{1}{2}\sum_q q^\dagger_+ \gamma_5 q_+ +
\sum_q q^\dagger_+\left({\vec r}\times i{\vec \partial}
\right)^z q_+  
+ \varepsilon^{+-ij}\mbox{Tr}F^{+i}A^j
+ 2 \mbox{Tr} F^{+j}\left({\vec r}\times i{\vec \partial} 
\right)^z A^j.
\label{M+12}
\ee
The first and third term in (\ref{eq:JJM}),(\ref{M+12}) are the
`intrinsic' contributions (no factor of ${\vec r}\times $) 
to the nucleon's angular momentum $J^z=+\frac{1}{2}$ and have a 
physical interpretation as quark and gluon spin respectively, while
the second and fourth term can be identified with the quark/gluon
OAM.
Here $q_+ \equiv \frac{1}{2} \gamma^-\gamma^+ q$ is the dynamical
component of the quark field operators, and light-cone gauge
$A^+\equiv A^0+A^z=0$ is implied. 
The residual gauge invariance can be fixed by
imposing anti-periodic boundary conditions 
${\bf A}_\perp({\bf x}_\perp,\infty^-)=-
{\bf A}_\perp({\bf x}_\perp,-\infty^-)$ on the transverse components
of the vector potential.

Only the $\Delta q$ are common to both decompositions. While for
a nucleon at rest the difference in the Dirac structure between
$L_q^z$ and ${\cal L}_q^z$ plays no role \cite{BC}, 
the appearance of the
gluon vector potential in the operator defining
$L_q^z$ implies that in general ${\cal L}_q^z\neq L_q^z$.
Other nucleon spin decompositions have been proposed in Refs.
\cite{Chen,Wakamatsu}.
\section{Quark OAM in the Scalar Diquark Model}
In the scalar diquark model \cite{spectator}, the light-cone
wave function for a `nucleon' with spin $\uparrow$ and the
quark spin aligned and anti-aligned respectively reads 
\cite{ELz}
\bea
\psi_{+\frac{1}{2}}^\uparrow \left(x,{\bf k}_\perp\right)
= \left(M+\frac{m}{x}\right) \phi (x,{\bf k}_\perp^2) 
\label{eq:SDQM}\quad \quad \quad \quad
\psi_{-\frac{1}{2}}^\uparrow (x,{\bf k}_\perp)
=-\frac{k^1+ik^2}{x} \phi (x,{\bf k}_\perp^2)
\eea
with $\phi = \frac{g/\sqrt{1-x}}{M^2-\frac{{\bf k}_\perp^2+m^2}{x}
-\frac{{\bf k}_\perp^2+\lambda^2}{1-x}}$.
Here $g$ is the Yukawa coupling and $M$/$m$/$\lambda$ are the masses 
of the `nucleon'/`quark'/diquark respectively. Furthermore
$x$ is the momentum 
fraction carried by the quark and ${\bf k}_\perp\equiv 
{\bf k}_{\perp e}-
{\bf k}_{\perp \gamma}$ represents the relative $\perp$ momentum.
Using these light-cone wave functions it is straightforward to
calculate a variety of observables that appear in the context
of nucleon spin physics. 
For example, since only $\psi^\uparrow_{-\frac{1}{2}}$ carries
(one positive) unit of OAM, 
the quark OAM according to JM is obtained as 
\be
{\cal L}_q = \int_0^1 dx \int \frac{d^2{\bf k}_\perp}{16\pi^3}
(1-x) \left|\psi_{-\frac{1}{2}}^\uparrow\right|^2,
\label{eq:LJM}
\ee where the factor $(1-x)$ is the fraction of 
OAM carried by the quark with momentum fraction $x$
in a two-body system with one unit of OAM \cite{BC}.
One can also use the above light-cone wave functions to calculate
the GPDs that enter the Ji relation, to calculate
\be
L_q \equiv \frac{1}{2}\!\int_0^1\!dx\, \left[xq(x)+xE(x,0,0)-
\Delta q(x)\right]. \label{eq:LJi}
\ee
Using a Lorentz invariant regulator, such as a Pauli-Villars 
regulator, one finds that the two definitions for quark OAM agree,
i.e. $L_q={\cal L}_q$ in the scalar diquark model, as was expected 
since $L_q^z$ in the scalar diquark model does
not contain a gauge field term.

\begin{figure}
\unitlength1cm
\begin{picture}(15,5.5)(4,14)
\includegraphics{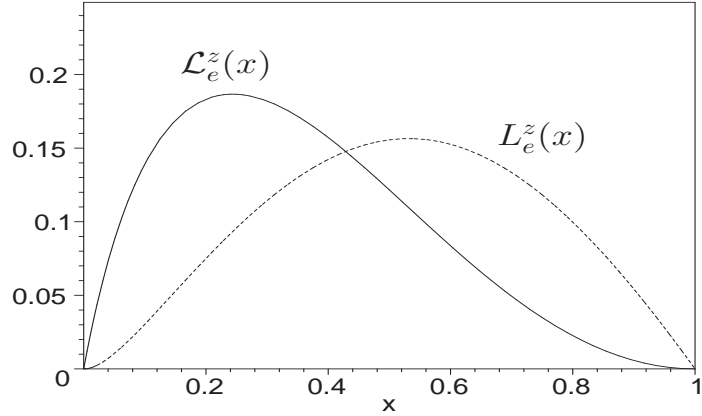}
\end{picture}
\caption{$x$ distribution of the `quark' OAM
${\cal L}_q^z(x)$ (full) compared to $L_q^z(x)$ from the unintegrated
Ji relation (dotted) in the scalar diquark model
for parameters $\Lambda^2=10m^2=10\lambda^2$.
Both in units of $\frac{g^2}{16\pi^2}$.} 
\label{fig:L(x)}
\centering
\end{figure}

However, despite $L_q={\cal L}_q$,
no such equality holds for the corresponding unintegrated
quantities. If one defines $L_q(x)$ by Eq. (\ref{eq:LJi})
without the $x$ integral \cite{hoodbhoy}, one does not obtain
the OAM distribution that would be obtained from (\ref{eq:LJM})
without $x$ integral (Fig.\ref{fig:L(x)}), which renders the interpretation
of $L_q(x)$ as the quark OAM distribution questionable.

\section{Electron OAM in QED}
In QED, there are four polarization states in the $e\gamma$
Fock component. To lowest order, the respective Fock space
amplitudes for a dressed electron with $J^z=+\frac{1}{2}$ read
\bea
\Psi^\uparrow_{+\frac{1}{2}+1}(x,{\bf k}_\perp)
&=& \frac{k^1-ik^2}{x(1-x)}\phi(x,{\bf k}_\perp^2)
\quad \quad \quad \quad
\Psi^\uparrow_{+\frac{1}{2}-1}(x,{\bf k}_\perp)
= -\frac{k^1+ik^2}{1-x}\phi(x,{\bf k}_\perp^2)\nonumber\\
\Psi^\uparrow_{-\frac{1}{2}+1}(x,{\bf k}_\perp)
&=& \left(\frac{m}{x}-m\right)\phi(x,{\bf k}_\perp^2) \quad \quad
\quad \quad
\Psi^\uparrow_{-\frac{1}{2}-1}(x,{\bf k}_\perp)=0
\eea
with $\phi(x,{\bf k}_\perp^2) = \frac{\sqrt{2}}{\sqrt{1-x}}\frac{e}{M^2-\frac{{\bf k}_\perp^2+m^2}{x}
-\frac{{\bf k}_\perp^2+\lambda^2}{1-x}}$. The label $\pm \frac{1}{2}$
represents the spin of the electron in the $e\gamma$ Fock component
and $\pm 1$ that of the $\gamma$.

Using these light-cone wave functions, it is again straightforward to
calculate the OAM 
of the electron in the JM \cite{JM} decomposition.
Including a Pauli-Villars subtraction with a heavy `photon' with mass $\Lambda$
one thus finds
\bea
{\cal L}_e^z = \int_0^1 \!\!dx \int\!\! \frac{d^2{\bf k}_\perp}{16\pi^3}
(1-x)\left[	
\left|\Psi^\uparrow_{+\frac{1}{2}-1}  \right|^2
\!\!-\left| \Psi^\uparrow_{+\frac{1}{2}+1}  \right|^2\right]
\stackrel{\stackrel{\Lambda \rightarrow \infty}{\lambda \rightarrow 0}}{\longrightarrow}-\frac{\alpha}{4\pi}\left[\frac{4}{3}\log \frac{\Lambda^2}{m^2}
-\frac{2}{9}\right]
\label{eq:JMQED}
\eea
Likewise, inserting the corresponding PDFs/GPDs from Ref. \cite{ELz}
into (\ref{eq:LJi}) yields
\be
L_e^z = \frac{1}{2}\int_0^1dx\, \left[x q_e(x)+xE_e(x,0,0)-
\Delta q_e(x)\right]\stackrel{\stackrel{\Lambda \rightarrow \infty}{\lambda \rightarrow 0}}{\longrightarrow}-\frac{\alpha}{4\pi}\left[
\frac{4}{3}\log \frac{\Lambda^2}{m^2}+\frac{7}{9}\right].
\label{eq:JiQED}
\ee
Both ${\cal L}_e^z$ and $L_e^z$ are negative, regardless of the
value of $\Lambda^2$ (as long as $\Lambda^2>\lambda^2$).
In the case of ${\cal L}_e^z$ the physical reason is a preference
for the emission of photons/gluons with the spin parallel 
(as compared to anti-parallel) to the original quark spin 
\cite{Ratcliffe} --- resulting more likely in a state with negative 
OAM. In fact, when $x\rightarrow 0$, this preference reflects the
more general principle of helicity retention \cite{BBS}, 
which favors the lead parton (i.e. the parton carrying most of the momentum) to carry a spin as close as possible
to that of the parent.  
It is also encoded in the evolution equations derived in Ref.
\cite{hoodbhoy}.
The divergent parts of ${\cal L}_e^z$ and $L_e^z$ are the same so 
that their difference is UV finite 
\be
{\cal L}_e^z - L_e^z \stackrel{\stackrel{\Lambda \rightarrow \infty}{\lambda \rightarrow 0}}{\longrightarrow} 
\frac{\alpha}{4\pi}. \label{eq:LLQED}
\ee

Applying these results to a (massive) quark with
$J^z=+ \frac{1}{2}$ yields to ${\cal O}(\alpha_s)$
\be
{\cal L}_q^z - L_q^z = \frac{\alpha_s}{3\pi}.
\label{eq:LLQCD}
\ee
This result may have important phenomenological consequences.
Recent lattice QCD calculations for GPDs yielded $L_u^z<0$ and
$L_d^z$>0, which is opposite to the sign obtained in typical
relativistic quark models --- such as the MIT bag model.

In QCD, the gluon spin is experimentally accessible, but the
gluon OAM ${\cal L}_g^z$ is not. On the other hand, the gluon
(total) angular momentum $J_g^z$ appearing in the Ji decomposition
is accessible, either indirectly (by subtraction, using quark GPDs), or directly, using gluon GPDs from lattice and/or deeply virtual
$J/\psi$ production. 
Even though $\frac{1}{2}\Delta G$ and $J_g^z$ belong to two
incommensurable decompositions of the nucleon spin, it is thus
tempting to consider the difference between these two quantities,
hoping to learn something about gluon OAM. Subtracting (\ref{eq:JJM})
from (\ref{eq:JJi}), it is straightforward to convince oneself that
$
J^z_g-\frac{1}{2}\Delta G = {\cal L}_g^z + 
\sum_q\left({\cal L}_q^z -L_q^z\right),
$
i.e. numerically $J^z_g-\frac{1}{2}\Delta G$ differs from 
${\cal L}_g^z$ by the same amount that $\sum_q{\cal L}_q^z$ differs 
from
$\sum_q L_q^z$. In our QED example, 
with 
\be
\Delta \gamma=\int_0^1 dx\int \frac{d^2{\bf k}_\perp}{16\pi^3}
\left[ \left|\psi_{+\frac{1}{2},+1}^\uparrow\right|^2
-\left|\psi_{+\frac{1}{2},-1}^\uparrow\right|^2 +
\left|\psi_{-\frac{1}{2},+1}^\uparrow\right|^2
-\left|\psi_{-\frac{1}{2},-1}^\uparrow\right|^2\right]
\ee
being the photon spin contribution, 
one thus finds (for $\lambda\rightarrow 0$, 
$\Lambda\rightarrow \infty$)
\be
J^z_\gamma-\frac{1}{2}\Delta \gamma = 
{\cal L}_\gamma^z + \frac{\alpha}{4\pi}.
\ee
As was the case in (\ref{eq:LLQED}), $\frac{\alpha}{4\pi}$ appears
to be a small correction, but one needs to keep in mind that  for
an electron
$J^z_\gamma$, $\Delta \gamma$, and ${\cal L}_\gamma^z$ are also
only of order $\alpha$.

\section{Quark OAM in the Bag Model}

Since quark models have, to lowest order in $\alpha_s$, no vector 
potential, it makes perhaps more sense to identify
the quark OAM from these models with ${\cal L}^z_q$ rather than with
the GPD-based $L^z_q$. In Ref. \cite{BC} the difference 
$\Delta^z_e \equiv {\cal L}_e^z-L_e^z$ was calculated to order
$\alpha$ for a single electron in QED and the result then also
applied to a single quark in QCD. However, in QCD quarks are never 
alone and the question arises regarding the effects from
`spectator currents' on the orbital angular momentum of each quark.

In order to address this issue,  we will
in the following use the MIT bag model \cite{MIT} to estimate ${\cal O}(\alpha_s)$
corrections to the difference
\be
\Delta^z_q \equiv L_q^z - {\cal L}_q^z= 
\langle q^\dagger\left({\vec r} \times g{\vec A}\right)^zq\rangle .
\label{eq:Delta}
\ee
The vector potential in (\ref{eq:Delta}) is calculated from
the spectator currents, which are obtained by taking matrix elements
in the corresponding ground state bag model wave functions.
The vector potential resulting from these static currents
is obtained by solving
\be
{\vec \nabla}^2 {\vec A}^{a}({\vec r}) = - {\vec j}^{a}({\vec r})
= - \sum_{s^\prime} g\psi^\dagger_{s^\prime}({\vec r})
{\vec \alpha} \frac{\lambda^a}{2} \psi_{s^\prime}({\vec r})
\ee
for each color component $a$ and
where the summation is over the spectators
(here we pick the gauge ${\vec \nabla}\cdot {\vec A}^a=0$, but 
to ${\cal O}(\alpha_s)$ the
result is actually gauge invariant --- at least
in the subclass of all gauges where all color components are
treated (globally) SU(3)-symmetrically. I such gauges, matrix elements
of operators of the type $q^\dagger \lambda^a \Gamma q A^a$,
where $\Gamma$ is some Dirac matrix, and $A^a$ is calculated 
to ${\cal O}(\alpha_s)$, are proportional to the
matrix elements of the corresponding abelian operators.
It is thus sufficient to establish gauge invariance of 
$q^\dagger {\vec r}\times {\vec A} q$ for abelian fields.
The key observation is that the bag model wave
functions contain no correlations between the positions of the 
quarks. Therefore, after eliminating the 
color in this calculation and introducing abelian currents, 
$\Delta_{s^\prime}^z$ factorizes into the
density of the active quark $\psi_s^\dagger({\vec r})\psi_s({\vec r})$
times $({\vec r}\times {\vec A})_z = rA_\phi$. 
Writing the volume integral $\int d^3r$ in cylindrical coordinates, 
one can isolate the only $\phi$-dependent term $r\int_0^{2\pi}d\phi
A_\phi= \oint d{\vec r}\cdot {\vec A}({\vec r})$ as a closed loop
integral with fixed $r$ and $z$. The closed loop integral is 
gauge invariant (its numerical value represents the color-magnetic
flux through a circle with radius $r$) and so is the volume
integral in which it enters.

The contribution from a spectator with $j^z=s^\prime$ to 
$\langle q ({\vec r}\times{\vec A})^z q \rangle$ thus
reads
\be
{\Delta}^z_{s^\prime} = -\frac{2}{3} \frac{g^2}{4\pi}
\int d^3r d^3r^\prime \psi_s^\dagger({\vec r})\psi_s({\vec r}) 
\psi_{s^\prime}^\dagger({\vec r}^\prime)
\frac{({\vec r}\times {\vec \alpha})^z}{|{\vec r}-{\vec r}^\prime|}
\psi_{s^\prime}({\vec r}^\prime) .
\ee
Note that 
${\vec \Delta}_{s^\prime}$ is independent of the angular momentum
$j^z=s$ of the
`active quark', since 
$\psi_{\frac{1}{2}}^\dagger({\vec r})\psi_{\frac{1}{2}}({\vec r})
=\psi_{-\frac{1}{2}}^\dagger({\vec r})\psi_{-\frac{1}{2}}({\vec r})$.
However, it depends on the spin of the spectator since the 
orientation of the vector potential entering (\ref{eq:Delta})
depends on the latter.
For example, for $s^\prime=+\frac{1}{2}$, one finds
\be 
\psi_{s^\prime}^\dagger({\vec r}^\prime)
{({\vec r}\times {\vec \alpha})^z}
\psi_{s^\prime}({\vec r}^\prime)
= |{\cal N}|^2 j_0(kr^\prime)j_1(kr^\prime)2\frac{xx^\prime +
yy^\prime}{r^\prime}
\ee
and hence independent of the bag radius \cite{Jarrah}
\be
{\Delta}^z_{\pm\frac{1}{2}}= \mp\frac{2}{3}\alpha_s |{\cal N}|^4\!\! \int_{r<R} \!\!\!\!\!\!\!\!d^3r \int_{r^\prime<R} \!\!\!\!\!\!\!\!d^3r^\prime \left[j_0^2(kr)+j_1^2(kr)\right]2\frac{xx^\prime+yy^\prime}{|{\vec r}-{\vec r}^\prime|}
\frac{j_0(kr^\prime)j_1(kr^\prime)}{r^\prime} = \mp0.78 \alpha_s.
\label{eq:integral}
\ee
When the active quark has $s$ aligned with that of the proton,
the two spectators must have opposite $s^\prime$ and their
contribution to ${\vec \Delta}$ cancels, i.e. ${\vec \Delta}$ is
nonzero only in those wave function components where the active quark
has $s=-\frac{1}{2}$, in which case both spectators have
$s^\prime=+\frac{1}{2}$. As a result, $\Delta^z_{q/p}$ is equal to
twice $\Delta^z_{+\frac{1}{2}}$ times the probability to find
that quark flavor with $s=-\frac{1}{2}$ (which is $\frac{1}{3}$ for
$q=u$ and $\frac{2}{3}$ for $q=d$, and hence
\bea
\label{eq:main}
\Delta_{u/p}^z = \frac{2}{3}{\Delta}^z_{+\frac{1}{2}}
=- 0.052\alpha_s\quad\quad\quad
\Delta_{d/p}^z = \frac{4}{3}{\Delta}^z_{+\frac{1}{2}}
=
- 0.104\alpha_s .
\eea


\begin{theacknowledgments}
This work was supported by the DOE under grant number 
DE-FG03-95ER40965.
\end{theacknowledgments}



\bibliographystyle{aipproc}   

\bibliography{sample}



\end{document}